\documentclass[12pt]{article}
\usepackage[a4paper, total={6.5in, 9.2in}]{geometry}
\usepackage[utf8]{inputenc}
\usepackage{graphicx}
\usepackage{algorithm}
\usepackage[noend]{algpseudocode}
\usepackage{amsmath} 
\usepackage{physics} 
\usepackage{xcolor} 
\usepackage{placeins} 
\usepackage{bm} 
\usepackage[T1]{fontenc} 
\usepackage{subcaption} 
\usepackage{hyperref}

\newcommand{\alg}[1]{\textsc{#1}}

\title{Subnetwork enumeration algorithms for multilayer networks}
\author{Tarmo Nurmi\thanks{Department of Computer Science, Aalto University School of Science, P.O. Box 15400, FI-00076, Finland} \thanks{tarmo.nurmi@aalto.fi} \and Mikko Kivel\"a\footnotemark[1]}
\date{}

\begin{document}

\maketitle

\begin{abstract}
To understand the structure of a network, it can be useful to break it down into its constituent pieces. This is the approach taken in a multitude of successful network analysis methods, such as motif analysis. These methods require one to enumerate or sample small connected subgraphs of a network, which can be computationally intractable if naive methods are used. Efficient algorithms exists for both enumeration and uniform sampling of subgraphs, and here we generalize the \alg{esu} algorithm for a very general notion of multilayer networks. We show that multilayer network subnetwork enumeration introduces nontrivial complications to the existing algorithm, and present two different generalized algorithms that preserve the desired features of unbiased sampling and trivial parallelization. We evaluate these algorithms in synthetic networks and with real-world data, and show that neither of the algorithms is strictly more efficient but rather the choice depends on the features of the data.  Having a general algorithm for finding subnetworks makes advanced multilayer network analysis possible, and enables researchers to apply a variety of methods to previously difficult-to-handle multilayer networks in a variety of domains and across many different types of multilayer networks.
\end{abstract}

\section{Introduction}

Efficiently finding connected, induced subgraphs is vital to many network analysis methods \cite{ribeiro2021survey}, from motif analysis \cite{milo2002network,wong2011biological,prvzulj2007biological,benson2016higher} to graph comparison with graph kernels \cite{pmlr-v5-shervashidze09a} and graph neural networks \cite{bouritsas2022improving}. These methods have found applications in a wide range of areas from cellular biology \cite{aittokallio2006graph,winterbach2013topology,babu2004structure} and neuroscience \cite{sporns2004motifs,morgan2018low} to software homology detection \cite{wu2018software}. In subgraph enumeration, the time to run the enumeration algorithm is often a bottleneck: The enumeration problem is highly combinatoric in nature, and real-world networks are often large enough that only small subgraphs can be practically enumerated, even with efficient algorithms \cite{wernicke2005faster,kashani2009kavosh,itzhack2007optimal,ribeiro2021survey}.
Even with efficient algorithms, in many cases the complete enumeration of all subgraphs of certain size is computationally infeasible, and uniform sampling of subgraphs is instead used to provide an estimate of the total occurrences. To this end, there are numerous algorithms for unbiased connected subgraph sampling \cite{wernicke2005faster,wernicke2006fanmod,ribeiro2010efficient,paredes2015rand}.
While most of the work on efficient graph enumeration is focused on networks represented with graphs at their simplest form, there have also been work on specific generalised network abstractions. For example, subnetwork enumeration algorithms have been developed for temporal networks \cite{boekhout2019efficiently,paranjape2017motifs} and multiplex networks \cite{takes2018multiplex,jiao2021sampling}, and theoretical bounds for subnetwork counting in combinations of graphs have been investigated \cite{enright2020dichotomies}. 

As more rich network data---e.g., networks with multiple modalities, different interactions, and/or several data dimensions---has become more and more common \cite{kivela2014multilayer,boccalletti2014structure,mcgee2019state,aldasoro2018multiplex,lee2020heterogeneous,buldu2018frequency,gligorijevic2015methods},  subnetwork enumeration for these types of networks has become an important task. Creating a custom subnetwork enumeration algorithm for each possible network type individually means that it is not applicable to the other network types.
This can be problematic as there is a multitude of different types of networks and slightly different variations of the most common network types, such as multiplex networks and temporal networks \cite{kivela2014multilayer}.  Further, the data can combine multiple network types: for example, one can have data on temporal networks that have multiple types of links. Developing algorithms for each specific network type would require one to solve technical difficulties that are very similar across the different types of networks. Here, instead, our aim is to find algorithms for general, application-agnostic subnetwork enumeration using a general multilayer network formalism that covers the above-mentioned network generalisations and many others \cite{kivela2014multilayer}.

We present two algorithms for enumeration and unbiased sampling of general multilayer connected induced subnetworks, \alg{nlse} and \alg{elsse}.
Both are generalizations of (\alg{rand-})\alg{esu} \cite{wernicke2005faster}, which is a subgraph enumeration algorithm based on recursively growing connected subgraphs by addition of neighboring nodes. The same recursive idea is applicable to multilayer networks, but instead of adding nodes as in \alg{esu}, the growth can be done via two different principles: \alg{nlse} adds a node-layer tuple and \alg{elsse} adds one node or an (elementary) layer at each step.
We show that both come with a compromise on how to only output unique connected subnetworks.
We compare their running times in random multilayer networks and in real-world networks, and show that their relative performance is dependent on network density, subnetwork size, and data-specific properties.

The algorithms presented are applicable to any multilayer networks with arbitrary intra- and interlayer edge structure, any number of aspects (dimensions), and possibly different sets of nodes participating on each layer. This enables one to analyse many different types of data with the two algorithms. Implementations of both algorithms are publicly available \cite{repository}. In total, our work provides an easy-to-apply tool for subnetwork enumeration in multilayer networks, and removes the need for algorithms developed individually for specific network types and their combinations.

\section{Preliminaries and definitions}

We start by describing the general form of multilayer networks and developing the concepts of multilayer subnetworks which cover a wide variety of real-world networked systems. Both are necessary preliminaries for section \ref{section:algorithms}, where multilayer subnetwork enumeration is presented.

\subsection{Multilayer networks}
\label{section:multilayer-networks}

Multilayer networks are networks where nodes exist on different layers and edges link nodes within and between the layers. Not all nodes are required to exist on all layers and there can be multiple dimensions of layers (called aspects). Multilayer networks are encountered in many networked systems where multiple modalities of interaction and different types of interacting actors are present \cite{kivela2014multilayer}. A multilayer representation is a more accurate description of such systems than a reduction to a network without layer information. For example, distance measures between multilayer networks based on the full multilayer information perform significantly better than those based on aggregating the network to a single layer (containing no meaningful layer information) \cite{sallmen2022graphlets}.

In a general multilayer network, the elements are instantiations of nodes within layers that have $d$ aspects to them, where $d$ corresponds to the "dimensionality" of the network \cite{kivela2014multilayer}. For example, in a two-aspect network of scientific citations, the nodes could be scientists, the first aspect could correspond to the year a scientist has been active in publishing, and the second aspect could correspond to the scientific discipline a scientist has been active in. The elements of the network are then scientist-year-discipline combinations, that is, a scientist $u$ exists on layer $(T,\alpha)$ where $T$ is a year and $\alpha$ is a scientific discipline. The tuple $(u,T,\alpha)$ is called a \emph{node-layer}, and the edges of the network are defined between node-layers. In this case, an edge $[(u,T,\alpha),(v,S,\beta)]$ would mean that there is a citation between the works of $u$ in year $T$ in discipline $\alpha$ and the works of $v$ in year $S$ in discipline $\beta$. Edges can be defined freely between node-layers: it is possible for a scientist to cite themselves from another year and same discipline, or a scientist to cite another's works from another year and discipline, or within the same discipline but different year, etc. Note that every node does not have to be present on every layer.

Formally, a multilayer network is defined as the quadruplet $M = (V_M,E_M,V,\mathbf{L})$ \cite{kivela2014multilayer}, where the layer information is encoded in the sequence of sets of \emph{elementary layers} $\mathbf{L} = \left(L_a\right)_{a=1}^d$. In the example above, the elementary layers of the first aspect would be years and the elementary layers of the second aspect would be scientific disciplines, that is, $L_1$ is the set of possible years and $L_2$ is the set of possible disciplines. In a general network, there are $d$ sets of elementary layers where $d$ is the number of aspects in the network. $V$ is the set of nodes (scientists in the example), and $V_M$ is the set of node-layers ($V_M \subseteq V \times L_1 \times L_2 \times ... \times L_d$). Edges $E_M$ are defined between node-layers, $E_M \subseteq V_M \times V_M$. Next, we define the notion of \emph{subnetworks} in multilayer networks.

\subsection{Multilayer subnetworks}
\label{section:multilayer_subnetworks}

To enumerate multilayer subnetworks, we need to define what a multilayer subnetwork is. More specifically, we are interested in \emph{induced subnetworks}, which are the subdivisions of the bigger network. 
Even more specifically, we are interested in \emph{connected} induced subnetworks, for reasons explained below. Connected induced subnetworks are the typical output of a subnetwork enumeration algorithm \cite{ribeiro2021survey}, and ours is no exception. Next, we define induced multilayer subnetworks, and then follow up with introducing subdivisions and connectedness, along with the definition of multilayer subnetwork size.

An induced subnetwork of a multilayer network is defined as a set of nodes and a set of elementary layers for each aspect, all node-layers within the span of those sets, and all edges between those node-layers \cite{sallmen2022graphlets}. In rigorous terms, an induced multilayer subnetwork $M'$ of a larger network $M = (V_M,E_M,V,\mathbf{L})$ is defined by $M' = (V'_M,E'_M,V',\mathbf{L}')$, where we have subset of nodes $V' \subseteq V$, subsets of elementary layers for each aspect $L'_a \subseteq L_a \; \forall \; a \in \{1,2,...,d\}$, subset of node-layer tuples  $V'_M = \{\bm{\gamma} \in V_M \; | \; \bm{\gamma} \in V' \times L'_1 \times L'_2 \times ... \times L'_d\}$, and subset of edges $E_M = \{ (\bm{\gamma},\bm{\delta}) \in E_M \; | \; (\bm{\gamma},\bm{\delta}) \in V'_M \times V'_M\}$. That is, $V'$ and $\mathbf{L}'$ characterize the span of the subnetwork in the space defined by elementary layers, and all the node-layers and edges in that span are included in the subnetwork. For example, in the citation network example in Section \ref{section:multilayer-networks}, a subnetwork would be defined by a set of scientists, a set of years, and a set of disciplines, all of which are subsets of those appearing in the original network. All network elements (node-layers and edges) appearing within the space defined by those sets are then included in the induced subnetwork.

Such an induced subnetwork is a qualitatively similar, smaller version of the larger multilayer network. This definition of subnetwork is useful as it preserves the inherent multilayer nature of the subnetwork, so that subnetworks as constituent pieces of a network retain the same dimensionality as the larger network. If we take all induced subnetworks of some fixed size and combine them together, we get the original network. In that way, the induced subnetworks are the subdivisions (building blocks) of the original network. That is why they are our main interest to enumerate.

However, enumerating \emph{all} subnetworks is not exactly what we want either. In subnetwork enumeration, the interest lies in \emph{connected} subnetworks, since any combination of any sets of elementary layers is technically a subnetwork (either connected or not connected). If we're looking for all subnetworks, connected or not, we have to go through all possible combinations of sets of elementary layers. In networks with low edge density, most of the subnetworks will be uninteresting because they contain no edges. Additionally, to find all subnetworks, there is no way to be more efficient than simple construction of all subsets and their combinations.
However, the set of connected subnetworks is a subset of the set of subnetworks. This is potentially a much smaller set, especially if the network is sparse (has low edge density). Being connected is often a property of interest, since it means for example information flow between nodes or node-layers.
In simple networks without layer information (single-layer networks, i.e. graphs), being connected means that there exists a path from every node to every other node. In multilayer networks, there are essentially two main definitions of what connectedness means, because different applications have had different requirements for example for information flow.

The first notion of connectivity for multilayer networks is that the \emph{underlying graph} (the ordinary graph formed by $V_M$ as nodes and $E_M$ as edges) is connected and the second is that the \emph{aggregated} network is connected \cite{sallmen2022graphlets}.
There are many ways to aggregate a multilayer network. However, regardless of the method of aggregation, the resulting aggregated network is either an ordinary graph (if all dimensionality was aggregated away) or another multilayer network.
The algorithms in this article are designed to enumerate multilayer subnetworks whose underlying graphs are connected. Together with existing algorithms to enumerate subgraphs, these are enough to cover both definitions of multilayer connectedness: in the case that the underlying graph should be connected, we can use the algorithms in this article, and in the case that the aggregated network should be connected, we can use the algorithms in this article on the aggregated network if it is a multilayer network or we can use a subgraph enumeration algorithm if it is an ordinary graph. Note that if we require the aggregated network to be connected and the aggregated network is a multilayer network, then we are again faced with the question of what it means for a multilayer network to be connected. The answer is either that the underlying graph should be connected or that the further-aggregated network of the aggregated network should be connected, at which point we're faced with the same question again. All aggregation methods therefore end up with either requiring the underlying graph to be connected or the aggregated network is an ordinary graph. It is therefore only necessary to construct a multilayer subnetwork enumeration algorithm that enumerates subnetworks whose underlying graphs are connected to cover both notions of connectivity.

Another relevant aspect of a multilayer subnetwork of interest is that it should not contain elementary layers that can be removed without affecting the node-layers and edges of the subnetwork. Essentially, this means that there should not be any elementary layers in any set $L'_a$ that are not found in at least one node-layer in $V'_M$. Equivalently, one should not be able to remove any elementary layer from any $L'_a$ without ending up with a different $V'_M$. If this criterion is met, we call the subnetwork \emph{minimal}. A non-minimal subnetwork essentially contains "ghost layers" without any nodes on them, or "ghost nodes" that don't exist on any layers. Such partial node-layers cannot have edges attached to them. In subnetwork enumeration, like connectedness, minimality is a desired property of the outputted subnetworks, since a non-minimal subnetwork is basically a redundant repetition of a minimal subnetwork (and non-minimal subnetworks can easily be constructed from minimal subnetworks). Therefore, we design the enumeration algorithms to output minimal connected induced subnetworks.

The size of a subnetwork is an important property in subnetwork-based network analysis methods. Often, we want to enumerate all subnetworks of a certain size. Therefore, the meaning of size of a multilayer subnetwork needs to be defined. A reasonable way is to consider the size of each of the sets of elementary layers, such that the size of subnetwork $M' = (V'_M,E'_M,V',\mathbf{L}')$ is $\abs{V'},\abs{L'_1},\abs{L'_2},...,\abs{L'_d}$, i.e. the size is characterized by $d+1$ numbers \cite{sallmen2022graphlets}. The algorithms presented in Section \ref{section:algorithms} are built to enumerate subnetworks whose size is defined as such.

\subsection{Subgraphs of underlying graphs}

An alternative way of defining a subunit of a multilayer network is to define it as a subgraph of the underlying graph of the multilayer network. This is different from the previous subnetwork definition, since a subgraph of the underlying graph isn't necessarily definable in terms of sets of elementary layers. For example, take a one-aspect network with node-layers $(1,\alpha),(2,\alpha),(2,\beta)$ where all of them are connected to each other, and take the subgraph formed by $(1,\alpha),(2,\beta)$; then, the subgraph cannot be defined as the sets of elementary layers $\{1,2\},\{\alpha,\beta\}$, since this span would also include the node-layer $(2,\alpha)$. We call subunits of a multilayer network defined by sets of node-layers \emph{subgraphs of underlying graphs}, to distinguish from the definition of multilayer subnetworks. Unlike subnetworks, subgraphs of underlying graphs are qualitatively different from multilayer networks as they are defined as subgraphs of $(V_M,E_M)$ rather than as quadruplets $M'=(V'_M,E'_M,V',\mathbf{L}')$. Enumeration of subgraphs of the underlying graph of a multilayer network does not require new algorithms; instead, existing subgraph enumeration algorithms can be directly applied to the underlying graph of the multilayer network.

\section{Multilayer subnetwork enumeration}
\label{section:algorithms}

\begin{figure}[!t]
    \centering
    \includegraphics[width=0.4\columnwidth]{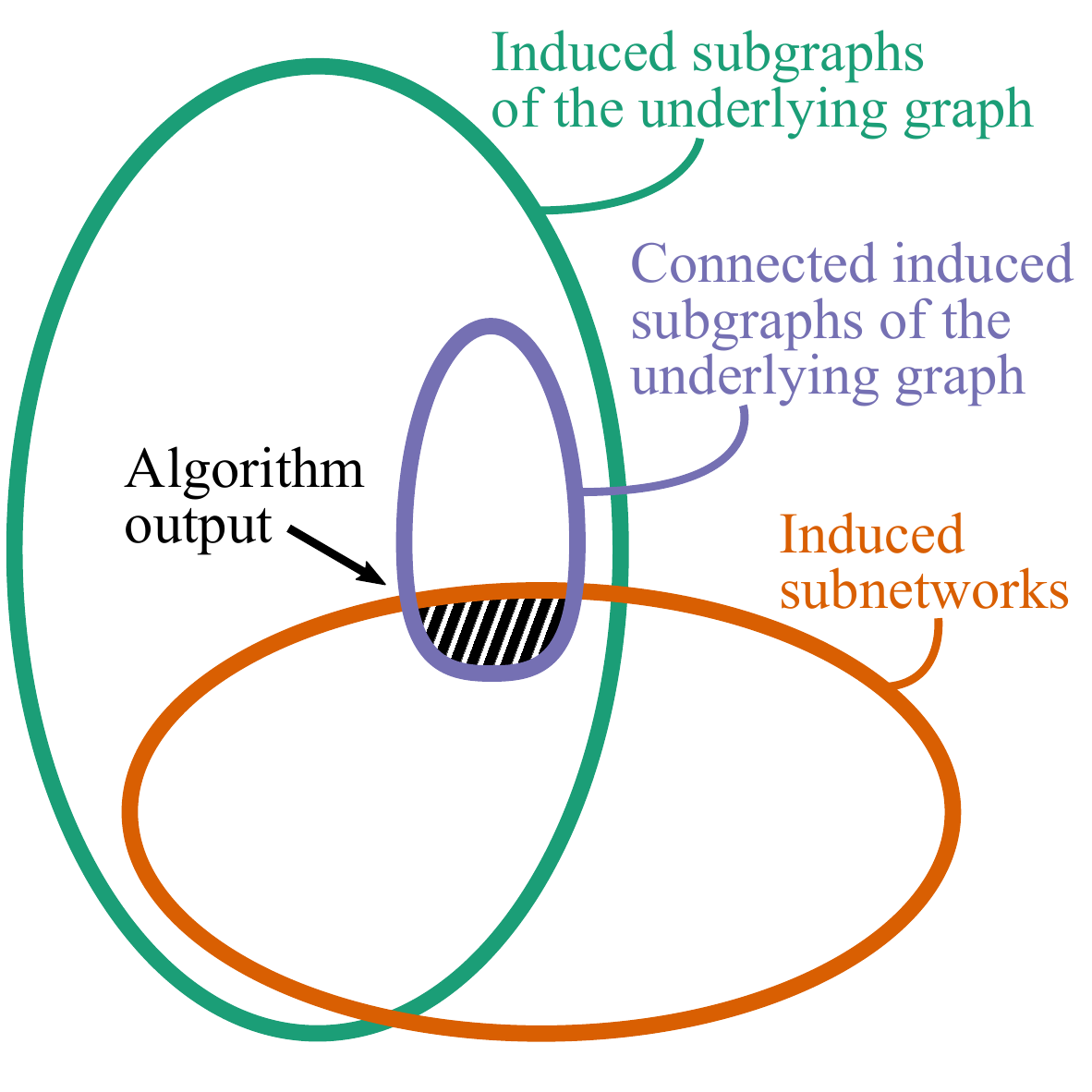}
    \caption{The relation of subgraphs and subnetworks in multilayer networks. The green set is the set of all possible induced subgraphs of the underlying graph of a multilayer network, and the orange set is the set of all possible induced subnetworks of the multilayer network. The intersection of the green and orange sets is the set of minimal induced subnetworks, i.e. subnetworks where every included elementary layer is found in at least one node-layer. The purple set is the set of connected induced subgraphs of the underlying graph, and it's a subset of the green set. The intersection of the purple and orange sets (shaded area) is the set of induced subnetworks which correspond to a connected induced subgraph of the underlying graph, in other words, the set of connected induced subnetworks. Connected induced subnetworks are the main interest in subnetwork enumeration, and therefore that set is also the output set of our algorithms.}
    \label{fig:subgraph_subnet_set}
\end{figure}

Figure \ref{fig:subgraph_subnet_set} illustrates the relationships between different subdivisions of multilayer networks, and the set of subnetworks that we want to enumerate (algorithm output). The enumeration problem can be approached from various angles, but many naive methods are highly inefficient (see Supplementary Information S1).
In sections \ref{section:nlse} and \ref{section:elsse}, we present two multilayer subnetwork enumeration algorithms, \alg{nlse} and \alg{elsse}. Both of them are based on recursive expansion of subnetworks, such that all subnetworks of a given size are explored and outputted exactly once. Outputting each subnetwork exactly once is an important property for unbiased subnetwork sampling -- in section \ref{section:sampling_and_parallelization}, we explain how to use \alg{nlse} and \alg{elsse} for sampling and how to parallelize their execution. In the single-layer setting, ensuring that each subnetwork is outputted exactly once is relatively straightforward, but in the multilayer case, additional checks are required (see sections \ref{section:nlse} and \ref{section:elsse}).

\subsection{\alg{nlse}}
\label{section:nlse}

\alg{node-layer set extension}, or \alg{nlse} for short, is presented in algorithm \ref{alg:nlse}. The input of the algorithm is a multilayer network and a size vector $\vb{s} = (s_0,s_1,s_2,...,s_d)$ where $s_i$ is the desired size of the set of elementary layers in aspect $i$ in the subnetworks. The output is all connected induced subnetworks of size $\vb{s}$.

\begin{algorithm}[!ht]
\small
\caption{\alg{nlse: node-layer set extension}}
\label{alg:nlse}
\vspace{1mm}
\textbf{Input:} Multilayer network $M = (V_M, E_M, V, \mathbf{L})$ with $d$ aspects, subnetwork size $\vb{s}$\\
\textbf{Output:} All connected induced subnetworks of size $\vb{s}$
\begin{algorithmic}[1]
\State Give each node-layer $\vb{\gamma} \in V_M$ a unique label $t_{\vb{\gamma}}$ by which they can be ordered
\For{each $\vb{\gamma} \in V_M$}
    \State $Extension \gets \emptyset$
	\For{each neighbor $\vb{\delta}$ of $\vb{\gamma}$}
	    \If{$t_{\vb{\delta}} > t_{\vb{\gamma}}$}
	        \State $Extension \gets Extension \cup \vb{\delta}$
        \EndIf
	\EndFor
	\State Call \alg{extend-nlse($M,\vb{s},\{\vb{\gamma}\},Extension,\vb{\gamma}$)}
\EndFor
\end{algorithmic}
\rule{\columnwidth}{0.5pt}
\alg{extend-nlse($M,\vb{s},V_{M,subnet},Extension,\vb{\gamma}$)}
\begin{algorithmic}[1]

\algrenewcommand{\alglinenumber}[1]{E#1:}

\If{$\abs{(\alg{SpannedVolume($V_{M,subnet}$)})_i} = s_i \; \forall \; i$}
	\If{\alg{Valid($M,V_{M,subnet},Extension,\vb{\gamma}$)}}
		\State \textbf{Output} \alg{SpannedVolume($V_{M,subnet}$)} and \textbf{return}
	\Else
		\State \textbf{return}
	\EndIf
\EndIf
\State $N \gets $ neighbors$(V_{M,subnet})$
\While{$Extension \neq \emptyset$}
    \State Pop random $\vb{\gamma}'$ from $Extension$
    \State \textbf{if} $\exists \; i: \abs{(\alg{SpannedVolume($V_{M,subnet} \cup \vb{\gamma}'$)})_i} > s_i$ \textbf{then goto} E7
    \State $Extension' \gets Extension$
    \For{each neighbor $\vb{\delta}$ of $\vb{\gamma}'$}
        \If{$t_{\vb{\delta}} > t_{\vb{\gamma}}$ and $\vb{\delta} \notin N$ and $\vb{\delta} \notin V_{M,subnet}$}
        \State $Extension' \gets Extension' \cup \vb{\delta}$
        \EndIf
    \EndFor
    \State Call \alg{extend-nlse($M,\vb{s},V_{M,subnet} \cup \vb{\gamma}',Extension',\vb{\gamma}$)}
\EndWhile
\State \textbf{return}
\end{algorithmic}

\rule{\columnwidth}{0.5pt}
\alg{SpannedVolume($V_{M,S}$)}

\textit{Return subnetwork $\mathbf{S}$ defined by node-layers in $V_{M,S}$}
\begin{algorithmic}[1]

\makeatletter
\def\ALG@step%
   {%
   \addtocounter{ALG@line}{1}%
   \addtocounter{ALG@rem}{1}%
   \ifthenelse{\equal{\arabic{ALG@rem}}{\ALG@numberfreq}}%
      {\setcounter{ALG@rem}{0}\alglinenumber{S\arabic{ALG@line}}}%
      {}%
   }%
\makeatother

\State \textbf{return} $\left( \{ \gamma_i \mid \vb{\gamma} \in V_{M,S} \} \right)_{i=0}^d$
\end{algorithmic}

\rule{\columnwidth}{0.5pt}
\alg{Valid($M,V_{M,S},Extension,\vb{\gamma}$)}

\begin{algorithmic}[0]

\makeatletter
\def\ALG@step%
   {%
   \addtocounter{ALG@line}{1}%
   \addtocounter{ALG@rem}{1}%
   \ifthenelse{\equal{\arabic{ALG@rem}}{\ALG@numberfreq}}%
      {\setcounter{ALG@rem}{0}\alglinenumber{V\arabic{ALG@line}}}%
      {}%
   }%
\makeatother

\State If the induced subnetwork defined by $\alg{SpannedVolume($V_{M,S}$)}$ is connected, and all node-layers in it have $t \geq t_{\vb{\gamma}}$, and $V_{M,S}$ doesn't have neighbors within \alg{SpannedVolume($V_{M,S}$)} that are not in $Extension$, \textbf{return true}. Otherwise, \textbf{return false}.
\end{algorithmic}
\end{algorithm}

The \alg{nlse} algorithm is a refined version of the approach of using the \alg{esu} \cite{wernicke2005faster} algorithm to enumerate the set of subgraphs of the underlying graph that contains the set of desired multilayer subnetworks.
The \alg{nlse} algorithm operates as \alg{esu} on the underlying graph of the network, and additionally it checks that the size of the subnetwork does not exceed $\vb{s}$ in any aspect while executing the algorithm. The idea is to find a subnetwork by discovering a \emph{spanning path} of node-layers inside a space defined in terms of elementary layers, not unlike a curve inside a higher-dimensional space.

The crucial element of the algorithm is the required validity check for a subnetwork of desired size. We must check that 1) the subnetwork is connected, 2) for all node-layers in the subnetwork, their label is greater or equal to that of the node-layer from which the current execution path was started, and 3) if there is a path already visited inside the subnetwork, evidenced by which node-layers are explored (removed from the list of possible future explorations) and which are not. These checks are not present in \alg{esu} operating on single-layer networks, but they are necessary for the multilayer case.

Next, we will go through the algorithm in detail.

\subsubsection{Detailed description}

\begin{figure}[t]
    \centering
    \includegraphics[width=\textwidth]{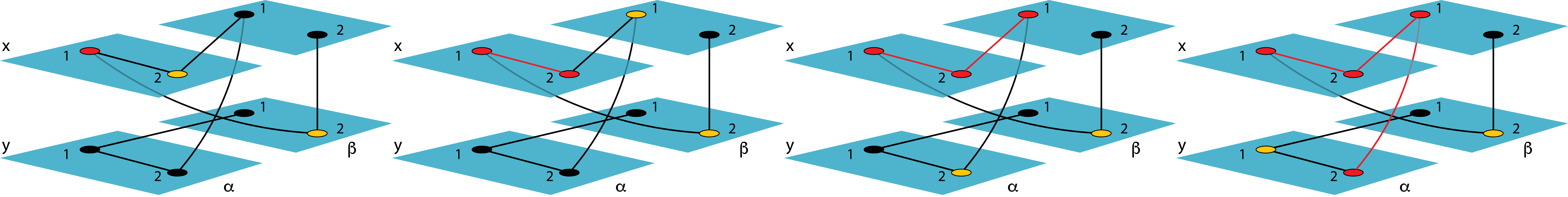}
    \caption{The operation of \alg{nlse}. Here, $(1,x,\alpha)$ is the starting node-layer $\vb{\gamma}$, which we assume to have the smallest label of the shown node-layers. The red node-layers are those that are in $V_{M,subnet}$ and the yellow ones are in $Extension$. The execution runs left to right, in recursive function calls to \alg{extend-nlse}. In the last image, the set of node-layers $V_{M,subnet}$ spans the whole subnetwork $\{1,2\},\{x,y\},\{\alpha,\beta\}$, and since all the neighbors of the red spanning path are in $Extension$ and the subnetwork is connected and all the labels of the node-layers in the subnetwork are larger than or equal to the label of the starting node-layer, the algorithm outputs the subnetwork.}
    \label{fig:nlse-example}
\end{figure}

The algorithm (algorithm \ref{alg:nlse}) is based on recursive exploration of subgraphs of the underlying graph. The function of the algorithm is illustrated in Figure \ref{fig:nlse-example}. First, each node-layer is given an unique label which provide an ordering for them (line 1). Then, for each node-layer, we start growing an execution tree from that node-layer (lines 2-7). We create an extension set, which holds the possible node-layers that are connected to the current subgraph of the underlying graph and can be added to it further up the execution tree (lines 3-6). Only node-layers with labels greater than the starting node-layer can be added to the extension set (line 5). The reason for the label comparison is to  ensure that a given subnetwork can only be discovered by starting the execution tree from the node-layer with the smallest label inside the subnetwork with the goal of preventing outputting a subnetwork multiple times.
The labeling is not enough to completely remove threat of duplicate output; additional checks are required when a potential subnetwork is discovered. The execution of the algorithm continues by calling the \alg{extend-nlse} subroutine (line 7), which then is recursively re-called within itself.

In \alg{extend-nlse}, the first thing done in each function call is to check if the space spanned by the current set of node-layers $V_{M,subnet}$ fulfills the size requirement $\vb{s}$ given for the multilayer subnetworks we want to enumerate (line E1). The set of node-layers $V_{M,subnet}$ forms a spanning path inside the space defined by the elementary layers of the multilayer network and the spanning path defines a subnetwork that contains that path. In a sense, the spanning path spans the subnetwork space. The spanned volume is then the actual subnetwork that we are looking to enumerate. The concept of spanning path is formalized in the subroutine \alg{SpannedVolume}, which takes as input a set of node-layers $V_{M,S}$ and outputs the subnetwork that the node-layer set spans. Simply put, for each aspect, the elementary layer set of the subnetwork for that aspect is the set of elementary layers of that aspect that are found in any of the node-layers of $V_{M,S}$. If the subnetwork outputted by \alg{SpannedVolume} that is spanned by $V_{M,subnet}$ matches the size requirement $\vb{s}$, the subnetwork is a potential candidate for a connected, induced subnetwork and the algorithm execution continues with the subroutine \alg{Valid} which checks if the subnetwork should be outputted (line E2).

The validity check \alg{Valid} takes as input the multilayer network $M$, the set of node-layers $V_{M,S}$ that define the subnetwork candidate, the node-layer extension set $Extension$, and the starting node-layer $\vb{\gamma}$. Since inside the spanned space of $V_{M,S}$ there might be isolated node-layers or larger disconnected subgraphs of the underlying graph which are invisible to the current execution path, \alg{Valid} checks that the subnetwork (not just the subgraph defined by $V_{M,S}$) is connected. A spanning path inside a subnetwork space is not necessarily unique, that is, there can be different sets of node-layers corresponding to the same spanned volume (the sets can be disjoint or overlapping). This could result in a subnetwork being outputted more than once, which we want to eliminate. To eliminate disjoint spanning paths, \alg{Valid} checks that all node-layers inside the subnetwork have their label larger than or equal to the label of the starting node-layer $\vb{\gamma}$ (equality only holds for $\vb{\gamma}$ itself). To eliminate overlapping spanning paths (i.e. different spanning paths that start from $\vb{\gamma}$), \alg{Valid} checks that the current spanning path $V_{M,S}$ does not have neighboring node-layers (node-layers that are connected to any node-layer in $V_{M,S}$) that are inside the subnetwork defined by \alg{SpannedVolume($V_{M,S}$)} but not in the extension set $Extension$. The presence of such neighboring node-layers would mean that there is already an explored spanning path that has resulted in outputting the subnetwork. In other words, only the branch of the overlapping spanning path that is the first one to be explored will result in outputting the subnetwork. The validity check \alg{Valid} therefore ensures that 1) only connected subnetworks are outputted and 2) each connected subnetwork is outputted exactly once.

If \alg{Valid} returns true, the subnetwork \alg{SpannedVolume($V_{M,subnet}$)} is outputted and algorithm execution returns to the previous function call (line E3). If the candidate subnetwork is not \alg{Valid}, the execution simply returns without outputting (lines E4-E5).

Once the check whether the current subgraph of the underlying graph $V_{M,subnet}$ is a valid subnetwork is done, the algorithm execution continues by constructing the set of neighbors of $V_{M,subnet}$ (set of neighbors of any node-layer in $V_{M,subnet}$, line E6). This is required to prevent the algorithm from repeatedly exploring already explored node-layers. Then, while there are node-layers in the extension set (line E7), we pop one arbitrarily chosen node-layer $\vb{\gamma}'$ from the set (line E8). First, we check if adding this node-layer to $V_{M,subnet}$ would make the subnetwork go over the size requirement (line E9), and if it does, we can skip the rest of the loop iteration. The execution continues by making a new extension set $Extension'$, which initially contains the same elements as $Extension$ (line E10), and to which we then add all the possible node-layers that the execution tree could be extended to after the addition of $\vb{\gamma}'$ to $V_{M,subnet}$ (lines E11-13). The requirements for adding a node-layer to $Extension'$ are that 1) they are neighbors of $\vb{\gamma}'$ (line E11) and 2) their labels are greater than that of the starting node-layer $\vb{\gamma}$, they are not in the neighborhood of $V_{M,subnet}$ ($V_{M,subnet}$ does not contain $\vb{\gamma}'$), and they are not already in $V_{M,subnet}$ (line E12).

After the new set of possible extension node-layers $Extension'$ has been constructed, the algorithm recursively calls \alg{extend-nlse} (line E14), using $V_{M,subnet} \cup \vb{\gamma}'$ as the new subgraph of the underlying graph (node-layer set) and $Extension'$ as the new set of possible node-layers to extend the subgraph into.

The algorithm thus finds spanning paths of possible multilayer subnetworks by checking possible extension directions and recursively exploring them, and upon finding a possible subnetwork checking if it is actually connected while making sure each subnetwork is only outputted once.

\subsection{\alg{elsse}}
\label{section:elsse}

\alg{Elementary layer set sequence extension}, or \alg{elsse}, is presented in algorithm \ref{alg:elsse}. It is based on recursive exploration of subspaces of the space defined by the elementary layers of the network, rather than exploration of spanning paths as in \alg{nlse}, aiming to more efficiently find multilayer subnetworks in situations where the spanning path approach is slow. Such situations could arise for example in higher-aspect networks (see section \ref{section:experimental}). The input is again a multilayer network and a size vector $\vb{s} = (s_0,s_1,s_2,...,s_d)$ where $s_i$ is the desired size of the set of elementary layers in aspect $i$ in the subnetworks. The output is all connected induced subnetworks of size $\vb{s}$.

\begin{algorithm}[!ht]
\small
\caption{\alg{elsse: elementary layer set sequence extension}}
\label{alg:elsse}
\vspace{1mm}
\textbf{Input:} Multilayer network $M = (V_M, E_M, V, \mathbf{L})$ with $d$ aspects, subnetwork size $\vb{s}$\\
\textbf{Output:} All connected induced subnetworks of size $\vb{s}$

\begin{algorithmic}[1]
\State Give each node-layer $\vb{\gamma} \in V_M$ a unique label $t_{\vb{\gamma}}$ by which they can be ordered
\For{each $\vb{\gamma} \in V_M$}
    \State $\mathbf{S} \gets ( \{\gamma_i\} )_{i=0}^d$
    \State $\mathbf{Extension} \gets ( \emptyset )_{i=0}^d$
	\For{each neighbor $\vb{\delta}$ of $\vb{\gamma}$}
	    \If{$t_{\vb{\lambda}} > t_{\vb{\gamma}} \; \forall \; \vb{\lambda} \in $ \alg{SubnetDiff($M,(S_i \cup \delta_i)_{i=0}^d,\mathbf{S}$)}}
	        \State $\forall \; i: Extension_i \gets Extension_i \cup \begin{cases} \delta_i & \textbf{ if } \delta_i \notin S_i\\ \emptyset & \textbf{ otherwise} \end{cases}$
        \EndIf
	\EndFor
	\State Call \alg{extend-elsse($M,\vb{s},\mathbf{S},\mathbf{Extension},\vb{\gamma}$)}
\EndFor
\end{algorithmic}

\rule{\columnwidth}{0.5pt}
\alg{extend-elsse($M,\vb{s},\mathbf{S},\mathbf{Extension},\vb{\gamma}$)}
\begin{algorithmic}[1]

\makeatletter
\def\ALG@step%
   {%
   \addtocounter{ALG@line}{1}%
   \addtocounter{ALG@rem}{1}%
   \ifthenelse{\equal{\arabic{ALG@rem}}{\ALG@numberfreq}}%
      {\setcounter{ALG@rem}{0}\alglinenumber{E\arabic{ALG@line}}}%
      {}%
   }%
\makeatother

\If{$\abs{S_i} = s_i \; \forall \; i$}
	\If{\alg{Valid($M,\mathbf{S}$)}}
		\State \textbf{Output $\mathbf{S}$} and \textbf{return}
	\Else
		\State \textbf{return}
	\EndIf
\EndIf
\State $\mathbf{N} \gets ( \{\nu_i \mid \vb{\nu} \in$ neighbors($\mathbf{S}$), $t_{\vb{\mu}} > t_{\vb{\gamma}} \; \forall \; \vb{\mu} \in$ \alg{SubnetDiff($M,(S_i \cup \nu_i)_{i=0}^d,\mathbf{S}$)} $\} )_{i=0}^d$
\While{$\exists \; j: Extension_j \neq \emptyset \text{ and } \abs{S_j} < s_j$}
    \State Choose a $j$ such that $Extension_j \neq \emptyset \text{ and } \abs{S_j} < s_j$
    \State Pop random $l$ from $Extension_j$
    \State $\mathbf{S'} \gets \left( S_i \cup \begin{cases} l & \textbf{ if } i = j\\ \emptyset & \textbf{ otherwise}\end{cases} \right)_{i=0}^d$
    \If{$t_{\vb{\beta}} \geq t_{\vb{\gamma}} \; \forall \; \vb{\beta} \in \left( \prod_{i=0}^d S'_i \right) \cap V_M$}
        \State $\mathbf{Extension'} \gets \mathbf{Extension}$
        \For{each $\vb{\tau} \in $ \alg{SubnetDiff($M,\mathbf{S'},\mathbf{S}$)}}
            \For{each neighbor $\vb{\delta} \notin \prod_{i=0}^d S'_i$ of $\vb{\tau}$}
                \If{$t_{\vb{\lambda}} > t_{\vb{\gamma}} \; \forall \; \vb{\lambda} \in $ \alg{SubnetDiff($M,(S'_i \cup \delta_i)_{i=0}^d,\mathbf{S'}$)}}
    	        \State $\forall \; i: Extension'_i \gets Extension'_i \cup \begin{cases} \delta_i & \textbf{ if } \delta_i \notin N_i, \delta_i \notin S'_i\\ \emptyset & \textbf{ otherwise} \end{cases}$
    	        \EndIf
            \EndFor
        \EndFor
        \State Call \alg{extend-elsse($M,\vb{s},\mathbf{S'},\mathbf{Extension'},\vb{\gamma}$)}
    \EndIf
\EndWhile
\State \textbf{return}
\end{algorithmic}

\rule{\columnwidth}{0.5pt}
\alg{SubnetDiff($M,\mathbf{S'},\mathbf{S}$)}

\textit{Return node-layers that are in subnetwork defined by $\mathbf{S'}$ but not $\mathbf{S}$}
\begin{algorithmic}[1]

\makeatletter
\def\ALG@step%
   {%
   \addtocounter{ALG@line}{1}%
   \addtocounter{ALG@rem}{1}%
   \ifthenelse{\equal{\arabic{ALG@rem}}{\ALG@numberfreq}}%
      {\setcounter{ALG@rem}{0}\alglinenumber{D\arabic{ALG@line}}}%
      {}%
   }%
\makeatother

\State \textbf{return} $\left( \prod_{i=0}^d S'_i \setminus \prod_{i=0}^d S_i \right) \cap V_M$
\end{algorithmic}

\rule{\columnwidth}{0.5pt}
\alg{Valid($M,\mathbf{S}$)}

\begin{algorithmic}[0]

\makeatletter
\def\ALG@step%
   {%
   \addtocounter{ALG@line}{1}%
   \addtocounter{ALG@rem}{1}%
   \ifthenelse{\equal{\arabic{ALG@rem}}{\ALG@numberfreq}}%
      {\setcounter{ALG@rem}{0}\alglinenumber{V\arabic{ALG@line}}}%
      {}%
   }%
\makeatother

\State If the induced subnetwork defined by $\mathbf{S}$ is connected and for every element in every $S_i$ there is at least one node-layer that contains that element, \textbf{return true}. Otherwise, \textbf{return false}.
\end{algorithmic}
\end{algorithm}

The procedure of \alg{elsse} upholds an extension set for each aspect of the network separately. In other words, it directly expands the space of the subnetworks, made of elementary layers, rather than a spanning path made of node-layers in its recursive step. In each step, one extension set (i.e. expansion direction, comparable to directions in Euclidean space) is chosen, and the subnetwork is extended in that direction. This extension brings a variable number of node-layers into the subnetwork, possibly even zero. Neighbors of the added node-layers are then checked for possibilities of expanding the subnetwork in their direction (i.e. adding their elementary layers into the extension sets). Eventually, the growing subnetwork will reach desired size as elementary layers are added to it.

Like \alg{nlse}, \alg{elsse} necessitates a validity check for a subnetwork candidate before it is outputted. Unlike \alg{nlse}, the check is very simple. We check that 1) the subnetwork is connected and that 2) it is minimal and thus contains no "empty space", that is, there are no elementary layers in the subnetwork which are not found in any actual node-layers in the subnetwork. The latter check is necessary because of the multi-dimensional (multiaspect) nature of multilayer networks. When expanding the extension sets, a neighboring node-layer can consist of elementary layers that are not found in the subnetwork in multiple different aspects. This means that adding one such elementary layer alone to the subnetwork might not add any actual node-layers into the subnetwork, since to add the neighboring node-layer, multiple elementary layers would have to be added. Since \alg{elsse} expands the subnetwork one elementary layer at a time, there might be steps in the execution at which no node-layers are added into the subnetwork. If there was no empty space check, a subnetwork could be outputted effectively multiple times, containing different "empty dimensions", and the outputted subnetworks might have wrong sizes. Most of the work associated with validity checking in \alg{nlse} (e.g. that by only starting from a specific node-layer we can output a subnetwork, ensuring each subnetwork is outputted only once) is done in \alg{elsse} when elementary layers are added into extension sets. Therefore, it is not straightforward to tell beforehand whether \alg{elsse} or \alg{nlse} is faster (see section \ref{section:experimental}).

Next, we will go through the algorithm in detail.

\subsubsection{Detailed description}

\begin{figure}[!t]
    \centering
    \includegraphics[width=\textwidth]{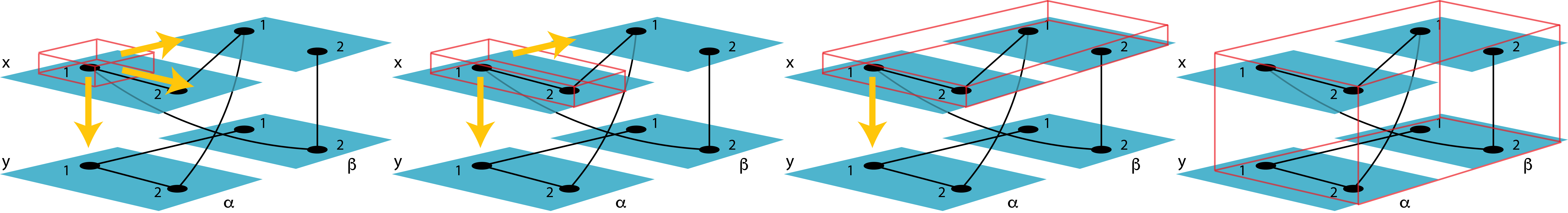}
    \caption{The operation of \alg{elsse}. Here, $(1,x,\alpha)$ is the starting node-layer $\vb{\gamma}$, which we assume to have the smallest label of the shown node-layers. The red box is the current subnetwork, i.e. the subspace defined in the space of the elementary layers. Initially it only encloses the starting node-layer. The yellow arrows show the extension possibilities in each aspect (there is a separate extension set for every aspect). In the first extension step, one node-layer $(2,x,\alpha)$ is added to the subnetwork. However, in the second extension step, the subspace span is extended to enclose two additional node-layers $(1,x,\beta)$ and $(2,x,\beta)$ in a single step, because the extension process expands the subspace directly and all combinations of existing elementary layers in other aspects with the newly added elementary layer are possible new node-layers. In the final step, four more node-layers are added, and the final subnetwork is $\{1,2\},\{x,y\},\{\alpha,\beta\}$. Since the subnetwork is connected and minimal (there is no "empty space", i.e. there is no elementary layer in any aspect that does not appear in any node-layer in the subnetwork), the algorithm outputs the subnetwork.}
    \label{fig:elsse-example}
\end{figure}

The algorithm (algorithm \ref{alg:elsse}) is based on recursive exploration of subspaces (i.e. subnetworks) of the space of elementary layers. The function of the algorithm is illustrated in Figure \ref{fig:elsse-example}. Each node-layer is given an unique label which provide an ordering (line 1), such that for a given subnetwork, there is only one node-layer from which the recursive process resulting in outputting that subnetwork can start. Initially, the execution starts from a single node-layer (lines 2-8), and the current subnetwork $\mathbf{S}$ is defined as a sequence of sets, one set of one elementary layer for each aspect, i.e. the elementary layers of the starting node-layer (line 3). Then, we initialize a sequence of extension sets, one for each aspect (line 4, initially each extension set is empty). These are the directions towards which the subnetwork can be extended. Then, we populate the extension sets by considering all neighboring node-layers of the starting node-layer (line 5), and seeing if all node-layers added to the subnetwork as a result of adding that neighboring node-layer's all elementary layers to $\mathbf{S}$ have labels greater than the starting node-layer's label (line 6). If yes, then the elementary layers of the neighboring node-layer are added to the extension lists (line 7), but only if they're not already in $\mathbf{S}$. The crucial function here is \alg{SubnetDiff}, which takes two subnetworks, $\mathbf{S'}$ and $\mathbf{S}$, and outputs the node-layers that are exclusively in $\mathbf{S'}$ but not in $\mathbf{S}$. In other words, if the difference between $\mathbf{S'}$ and $\mathbf{S}$ is that $\mathbf{S'}$ contains all elementary layers of some specific node-layer in addition to all the elementary layers in $\mathbf{S}$, then their \alg{SubnetDiff} returns all the node-layers that would be added to $\mathbf{S}$ as a result of adding that specific node-layer's elementary layers to $\mathbf{S}$. The recursive extension function \alg{extend-elsse} is then called with the constructed $\mathbf{S}$ and extension sets, and the original starting node-layer $\vb{\gamma}$ (line 8).

\alg{extend-elsse} begins with checking whether the current subnetwork candidate $\mathbf{S}$ is a valid one (lines E1-E5). It has to fulfill the size requirement $\mathbf{s}$ (line E1) and the validity check \alg{Valid} (line E2). Since it is possible to arrive at a connected subnetwork via a disconnected intermediary subnetwork, we have to check that the subnetwork is connected. In other words, a growing intermediary subnetwork that is disconnected at some point of the execution of the algorithm, may or may not eventually become connected when reaching the desired subnetwork size $\vb{s}$. On the other hand, some connected subnetworks can only be reached via disconnected intermediary subnetworks. Thus we can't know if a disconnected intermediary subnetwork will yield a connected subnetwork, and we also have to explore the disconnected intermediary subnetworks to find all connected subnetworks. Combined, this means that we have to check for connectedness upon finding a subnetwork candidate.

We also have to check that there are no "ghost" elementary layers in the subnetwork, i.e. that there are no elementary layers that are not found in any of the node-layers of $\mathbf{S}$. In other words, the subnetwork defined by $\mathbf{S}$ should be minimal, i.e. no elementary layers can be removed from any aspect without removing at least one actual node-layer from the subnetwork (see section \ref{section:multilayer_subnetworks}). Having an extension list separately for every aspect can result in the addition of "ghost" elementary layers into the subnetwork during the course of the algorithm execution, and since that can serve as a necessary intermediate step we cannot and should not prevent such additions. Instead, since the final output should only contain minimal subnetworks, we check for minimality before outputting. We only want to output minimal subnetworks because a non-minimal subnetwork has the same node-layer-and-edge structure as some minimal subnetwork and thus outputting non-minimal subnetworks in the algorithm would result in outputting essentially the same subnetworks unpredictably many times, over and over again.

If the subnetwork candidate is not found to fulfill the size requirement yet, we move to the rest of \alg{extend-elsse}. Similar to \alg{esu} and \alg{nlse}, we calculate the "original neighborhood" $\mathbf{N}$ of the current subnetwork to avoid exploring subnetworks multiple times further up the execution tree (line E6). This is done by finding all neighbors of $\mathbf{S}$, that is, all node-layers that are connected to any node-layer in the subnetwork defined by $\mathbf{S}$ and that are not contained in the subnetwork itself. Then, for every such neighbor, if adding that neighbor's elementary layers to $\mathbf{S}$ would only add node-layers whose labels are greater than the starting node-layer's label, the neighbor's elementary layers are added to the original neighborhood $\mathbf{N}$. $\mathbf{N}$ itself is a sequence, and by construction defines those elementary layers in every aspect that were added to the extension sets in the past (in previous function calls closer to the root in the recursion tree). This can be seen from the steps where elements are added to the next function call's extension sets (lines E14-E16). $\mathbf{N}$ defines the forbidden directions into which the subnetwork is not allowed to be extended.

In the main loop of \alg{extend-elsse}, we choose one of the nonempty extension sets corresponding to an aspect in which the subnetwork does not fulfill the size requirement yet, and the loop is continued while there is at least one such set that can be chosen (lines E7-E17). The subnetwork is then extended into that direction by removing one elementary layer $l$ from the chosen set (lines E8-E9), and constructing a new subnetwork $\mathbf{S}'$ as a combination of $\mathbf{S}$ and the elementary layer $l$ (line E10). Since this extension operation can add one or more node-layers into the subnetwork, we check that the subnetwork node-layers have labels at least as great as the starting node-layer $\vb{\gamma}$ (line E11, equality only holds for $\vb{\gamma}$ itself). This is done to eliminate multiple outputs of the same subnetwork by ensuring that there is only exactly one node-layer from which the recursive process for outputting the subnetwork (path from the root to a leaf of the recursion tree) can start. This check is necessary since when adding elementary layers to the extension sets later in the function execution, we check that individually their addition does not conflict with the label requirement for introduced node-layers (line E15), but \emph{combinations} of adding elementary layers to the subnetwork space might add node-layers whose labels are less than the starting node-layers. Rather than go through all possible combinations and keeping track of which combinations would not be allowed at the time of adding elementary layers to the extension sets, we do the checking after extending the subnetwork and before the costly operation of finding the new valid extension directions. The check at line E11 could thus be done at various points of the algorithm in slightly different ways, but making the check here makes sense since no extra bookkeeping with combinations of the extension directions is necessary this way.

Once the subnetwork space $\mathbf{S}$ has been extended into a direction, creating $\mathbf{S}'$, we check what future extension directions this extension opens up (lines E12-E16). First, we construct $\mathbf{Extension}'$ to which we copy the currently existing extension sequence (line E12). Then, $\mathbf{Extension}'$ is augmented by additional extension possibilities that can be explored further down the recursion tree. For every node-layer that was added in the extension process (line E13), we check every one of its neighbors that's not already in the extended subnetwork space $\mathbf{S}'$ (line E14), and for every such neighbor, we check whether adding that neighbor's elementary layers into $\mathbf{S}'$ would only introduce node-layers whose labels are greater than the starting node-layer's label (line E15). If so, then the neighbor's elementary layers can be added into their respective aspects' $Extension'$ sets, provided that the elementary layer is not in the "forbidden directions" set in sequence $\mathbf{N}$ or already included in $\mathbf{S}'$ (line E16).

Once the new extension set sequence has been constructed, the algorithm recurses and calls \alg{extend-elsse} with the new subnetwork space $\mathbf{S}'$ and extension set sequence $\mathbf{Extension}'$ (line E17).

The algorithm therefore finds subnetworks by directly expanding the subnetwork space (rather than e.g. following spanning paths of node-layers) in a recursive fashion while ensuring that each subnetwork is outputted exactly once.

\subsection{Sampling and parallelization}
\label{section:sampling_and_parallelization}

The algorithms \alg{nlse} and \alg{elsse} can be transformed into versions that output an unbiased sample of subnetworks instead of enumerating all subnetworks and/or can be parallelized, with little modification. Both are important features, since the computation time required to enumerate all subnetworks can be substantial due to large number of subnetworks. Sampling reduces the required overall computation time and parallelization reduces the required elapsed real-world wall clock time by distributing the computation effort onto many computational units. In this section, we describe how the algorithms are to be modified for sampling and parallelization.

\subsubsection{Sampling}

The recursive execution of the algorithms \alg{nlse} and \alg{elsse} creates a recursion tree whose leaves are the subnetworks of desired size. The algorithms can easily be transformed into unbiased subnetwork sampling algorithms by applying a probability of exploring a branch instead of always exploring all branches. Only a sample of branches are then explored all the way to the leaf, thus only a sample of subnetworks is outputted. If every leaf has \emph{a priori} the same probability of being explored, the sample is unbiased. Next, we describe a branch exploration strategy that has that property for both algorithms.

For both \alg{nlse} and \alg{elsse}, the exploration of a subnetwork starts from an initial node-layer (line 2 in algorithms \ref{alg:nlse} and \ref{alg:elsse}). That is the start of a path to one or more leaves on the recursion tree. We start exploring towards the leaves with probability $p_0$, that is, we completely skip the subnetwork exploration starting from that node-layer with probability $1-p_0$. The leaf at the end of a path corresponds to a subnetwork of size $\vb{s} = (s_0,s_1,s_2,...,s_d)$, where $d$ is the number of aspects. Therefore, during the course of exploration of a path in the execution tree, $(s_0-1)+(s_1-1)+(s_2-1)+...+(s_d-1)$ elementary layers are added to the subnetwork (the execution path starts from a node-layer which already has one elementary layer in every aspect). By applying a probability $p_i$ of continuing the execution whenever the $i$th elementary layer is added to the subnetwork we can sample a subnetwork with a probability equal to the product of all the applied probabilities (including $p_0$). In other words, we define a vector of probabilities $\vb{p} = (p_0,p_1,p_2,...,p_k)$, where $k = \sum_{i=0}^d (s_i - 1)$, and the probability of sampling a subnetwork is $\prod_{i=0}^k p_i$.

The vector $\vb{p}$ can be defined the same way for both \alg{nlse} and \alg{elsse}. However, the specific modifications of the algorithms are slightly different, since in \alg{elsse} we always add one elementary layer at a time to the growing subnetwork and in \alg{nlse} we can add multiple elementary layers at a time. In the latter case, we need to carefully make sure that all probabilities $p_i$ are applied to guarantee that every subnetwork is sampled with the same probability. If the actual subnetwork, defined by \alg{SpannedVolume}, in \alg{nlse} is not increased in any aspect at all, we continue algorithm execution with probability 1.

In order to transform the algorithms into sampling versions, there are two modifications that should be applied to both algorithms, and one modification that is different between the two algorithms.
For both algorithms (\ref{alg:nlse} and \ref{alg:elsse}), we 1) add vector of probabilities $\vb{p}$ to requirements and 2) at the end of line 2 in each algorithm, add "with probability $p_0$". For \alg{nlse} (algorithm \ref{alg:nlse}), after line E8 and before line E9, we add a line  "with probability $\prod_{i=j}^{j'} p_i$ execute lines E9-E14", where
\begin{align*}
    j &= 1 + \sum_{i=0}^d (\abs{S_i}-1) \\
    j' &= (j - 1) + \sum_{i=0}^d (\abs{S'_i}-\abs{S_i}) \\
    \mathbf{S} &= \alg{SpannedVolume($V_{M,subnet}$)} \\
    \mathbf{S'} &= \alg{SpannedVolume($V_{M,subnet} \cup \vb{\gamma'}$)}.
\end{align*}
Note that if $j' < j$, the product $\prod_{i=j}^{j'} p_i$ is empty and by convention has value 1. This corresponds to the case where the addition of a node-layer $\vb{\gamma'}$ into $V_{M,subnet}$ doesn't expand the spanned volume ($\mathbf{S'} = \mathbf{S}$).
For \alg{elsse} (algorithm \ref{alg:elsse}), after line E9 and before line E10, we add a line "with probability $p_i$ execute lines E10-E17", where $i = 1 + \sum_{i=0}^d (\abs{S_i}-1)$.

The sampling is unbiased in the sense that \emph{a priori} every subnetwork has the same probability of being explored, namely, the product of the probabilities of exploring a branch up to the leaf. However, since the execution paths for different subnetworks are not in general independent from each other (subnetworks may share elements), the individual sampled subnetworks are not independent.

\subsubsection{Parallelization}

In both algorithms \ref{alg:nlse} and \ref{alg:elsse}, the main loop runs over the node-layers (line 2 in both algorithms, "\textbf{for} each $\vb{\gamma} \in V_M$ \textbf{do}"). Each iteration of the loop is independent, except for the initial labeling of the node-layers (line 1 in both algorithms). Therefore, the algorithms are immediately parallelizable by first labeling the node-layers, and then running a parallel execution of the iterations of the main loop. In other words, instead of running the loop on line 2 sequentially, each iteration can be run independently in parallel, making the elapsed real time of the algorithm execution the same as the most time-consuming single iteration of the loop.

\section{Experimental evaluation}
\label{section:experimental}

We evaluate the absolute and relative performances of the two algorithms \alg{nlse} and \alg{elsse} via an implementation in C++ (available at \cite{repository}). We investigate the performance and scaling behavior 1) in a real-world single-aspect multiplex network set (protein-protein interaction networks), 2) in networks created from two single-aspect multiplex random network models, and 3) in networks created from a multi-aspect multilayer random network model. Multiplex networks are an important subtype of multilayer networks, where edges between layers are only from a node on a layer to the same node on another layer (within layers edges can exist between any nodes), and they are often encountered when dealing with real-world networks. The evaluation was done on off-the-shelf hardware (Intel Xeon CPU E5-2620 v3, 2.40GHz, introduced in 2014) in linear execution as a single process with no parallelization.

\subsection{Real-world single-aspect multiplex networks}
\begin{figure}[!t]
    \centering
    \begin{subfigure}[t]{0.6\textwidth}
        \vskip 0pt
        \includegraphics[width=\textwidth]{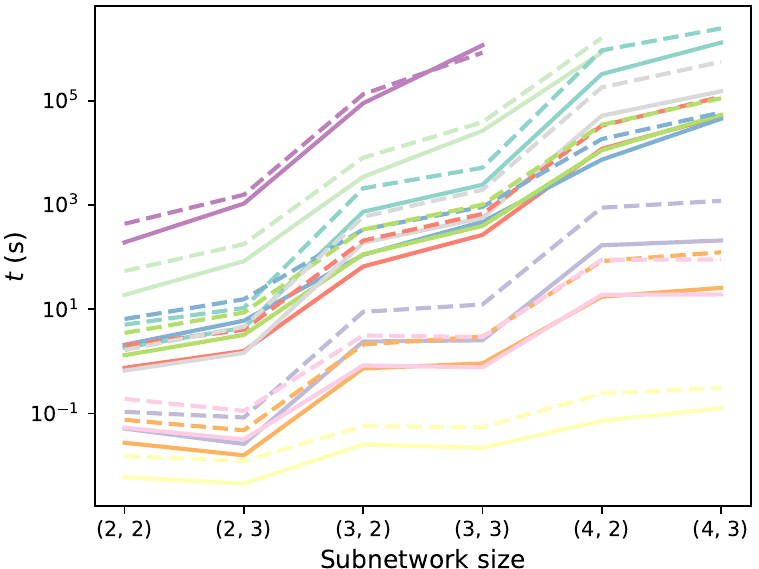}
    \end{subfigure}
    \hspace{0.05\textwidth}
    \begin{subfigure}[t]{0.17\textwidth}
        \vskip 0pt
        \includegraphics[width=\textwidth]{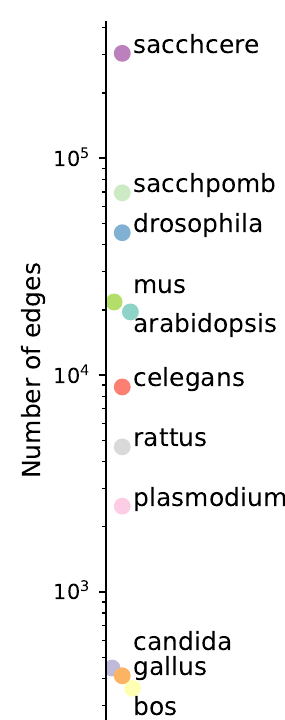}
    \end{subfigure}
    \caption{Running times on multiplex protein-protein interaction networks as function of subnetwork size. The solid line is \alg{nlse} and the dashed line is \alg{elsse}. The number of edges in each network is shown on the right, and full information on network size is presented in Table \ref{table:ppi_net_sizes}.}
    \label{fig:cpp_relative}
\end{figure}

We use multiplex protein-protein interaction networks \cite{de2015structural,stark2006biogrid} of varying sizes from diverse organisms in our evaluation as example of real-world data, with algorithm running times shown in Figure \ref{fig:cpp_relative}. The algorithms' running times scale sharply with increasing subnetwork size, naturally, since the number of (possible) subnetworks increases very quickly with size. Additionally, the size of the network has a great effect on the running time (different organisms have different numbers of nodes, layers and edges in their networks, shown in Table \ref{table:ppi_net_sizes}).
It seems that increasing the number of layers by one in the subnetworks has less of an effect on the running time than increasing the number of nodes by one, since the number of layers in the networks is smaller than the number of nodes.
In almost all the organisms and subnetwork sizes, \alg{nlse} finds the subnetworks faster, with the exception of subnetwork size $(3,3)$ in organism sacchcere (\textit{Saccharomyces cerevisiae}).

\begin{table}[!b]
    \centering
    \caption{Sizes of the protein-protein interaction networks.}
    \begin{tabular}{c|ccc}
         & edges & nodes & layers \\ \hline
         arabidopsis & 19 574 & 6 980 & 7 \\
         bos & 360 & 325 & 4 \\
         candida & 446 & 367 & 7 \\
         celegans & 8 826 & 3 879 & 6 \\
         drosophila & 45 401 & 8 215 & 7 \\
         gallus & 411 & 313 & 6 \\
         mus & 21 753 & 7 747 & 7 \\
         plasmodium & 2 489 & 1 203 & 3 \\
         rattus & 4 670 & 2 640 & 6 \\
         sacchcere & 304 886 & 6 570 & 7 \\
         sacchpomb & 69 324 & 4 092 & 7 
    \end{tabular}
    \label{table:ppi_net_sizes}
\end{table}

The numbers of found subnetworks, along with algorithm running times and output rates, are shown in Supplementary Information S2.
The largest enumeration that was still enumerable in reasonable time was 3.48 billion subnetworks for subnetwork size 4,2 in organism sacchpomb. In general, the output rates for the algorithms varied mostly between $10^2$ to $10^4$ subnetworks/second. The absolute rate is naturally dependent on hardware: the aforementioned subnetworks of size 4,2 in sacchpomb were enumerated at rate $4.16 \times 10^3$ subnetworks/second on the main testing hardware and at rate $1.17 \times 10^4$ subnetworks/second on the more powerful Intel Core i5-12600K desktop CPU, using the \alg{nlse} algorithm.

\subsection{Synthetic single-aspect multiplex networks}
For evaluation on synthetic multiplex random network models, we use two models: multiplex Er\H{o}s-Rényi model and multiplex geometric random network model. In the Erd\H{o}s-Rényi model, each possible multiplex network with a fixed number of intralayer edges per layer has the same probability of being realized (drawn from the ensemble), and in the geometric model, the intralayer networks are soft random geometric graphs with approximately a fixed number of edges per layer and where node positions are the same on all layers. In both models, interlayer-wise all nodes are connected to their counterparts on the other layers. The Erd\H{o}s-Rényi model corresponds to a situation where there is no specific structure in the network and no particular edge overlaps or correlations between layers, and the geometric model corresponds to a situation of high edge overlap between layers. In both models, we have 1000 nodes and set the number of edges within each layer to 1500 (mean intralayer degree 3 within each layer), and increase the number of layers from 3 to 20. 

\begin{figure}[!t]
    \centering
    \includegraphics[width=0.6\columnwidth]{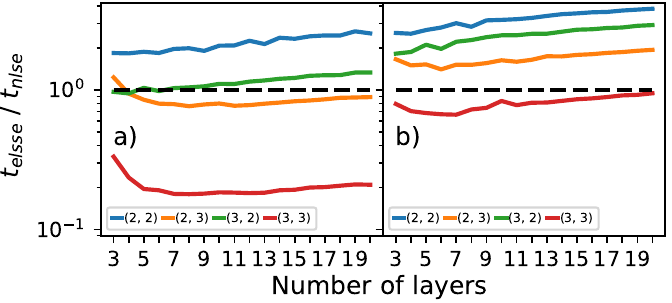}
    \caption{Relative running times as function of number of layers in a) multiplex geometric random networks with the same node locations on every layer and b) multiplex Er\H{o}s-Rényi networks. Every layer had 1000 nodes and 1500 edges (average intralayer degree 3). Colors correspond to subnetwork sizes. Above the dashed line \alg{nlse} is faster, and below \alg{elsse} is faster.}
    \label{fig:cpp_mplex}
\end{figure}

The relative running times for various subnetwork sizes are shown in Figure \ref{fig:cpp_mplex}. We see that the bigger the subnetworks to be enumerated are, the better \alg{elsse} performs compared to \alg{nlse}. 
The point where \alg{elsse} becomes faster than \alg{nlse} with respect to subnetwork size depends on the network model, i.e. the intralayer structure and edge overlaps between layers of the networks. With the geometric model, the relative running times are shifted in favor of \alg{elsse} as compared to the Erd\H{o}s-Rényi model, showing that when there is edge correlation structure between layers, the transition point where \alg{elsse} becomes faster than \alg{nlse} occurs at smaller subnetwork size.
The difference in running times can be quite significant: \alg{elsse} is at best about 5 times faster than \alg{nlse} for large subnetworks in the geometric model case and \alg{nlse} is at best about 4 times faster for small subnetworks in the Erd\H{o}s-Rényi model case. In general, it seems that for larger subnetwork sizes \alg{elsse} should be used. Additionally, for both models, the general tendency seems to be that increasing the number of layers in the network increases the ratio of \alg{elsse}'s and \alg{nlse}'s running times, such that for a large number of layers, \alg{nlse} might be faster for larger subnetworks as well.

\subsection{Synthetic multi-aspect multilayer networks}
In order to investigate the relative performance on higher-aspect networks with interesting multilayer structure also \emph{between} layers, we construct two-aspect geometric random multilayer model networks. In these networks, the intralayer and the interlayer networks are all soft random geometric graphs with the same node locations. Therefore, there are connections from nodes on layers to other, physically close nodes on different layers, creating a complex multilayer structure.

\begin{figure}[!t]
    \centering
    \includegraphics[width=0.6\columnwidth]{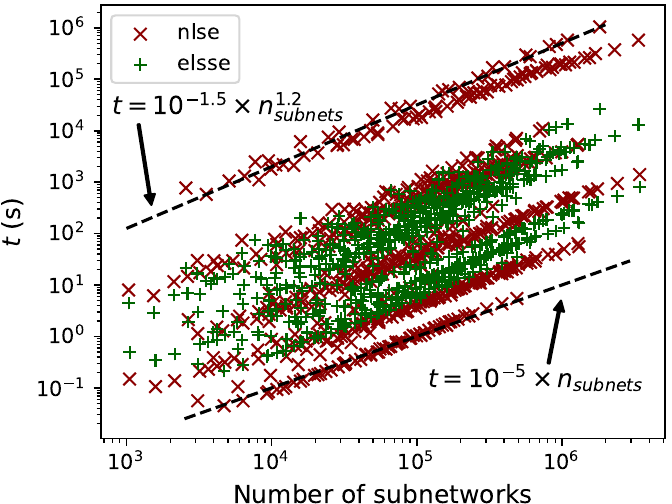}
    \caption{\alg{nlse} and \alg{elsse} running times versus the number of subnetworks found for two-aspect multilayer geometric random networks. The network sizes varied from 1000 to 10 000 nodes (1000-5000 in steps of 500, then jumping to 10 000), 3 to 10 layers in the first aspect, and 1 to 2 layers in the second aspect (1 to 4 in 1000-node networks). The subnetwork sizes varied from 2,2,1 to 3,3,2 in the same combinations depending on the number of elementary layers in the second aspect as in Figure \ref{fig:cpp_geo_mlayer}. The network size and subnetwork size strongly affect the number of subnetworks that a network contains. With the variations in different aspects and network sizes, we see that in general the algorithm execution times scale linearly as a function of the number of subnetworks in log-log space. The black dashed lines show curves with equations $t = 10^{-1.5} \times n_{subnets}^{1.2}$ and $t = 10^{-5} \times n_{subnets}$, where $n_{subnets}$ is the number of subnetworks. The scaling behaviors (i.e. exponent $k$ of $t = constant \times n^k$) of the algorithms lie between these two curves, and thus the algorithms have time complexities somewhere between $\mathcal{O}(n)$ and $\mathcal{O}(n^{1.2})$, where $n$ is the number of subnetworks.}
    \label{fig:cpp_geo_absolute_scatter}
\end{figure}

We create networks with a variable number of nodes (1000-5000 in steps of 500, and we also create networks with 10 000 nodes), fixed intralayer degrees (mean intralayer degree of 3, i.e. number of intralayer edges per layer ranging from 1500 to 15 000 as number of nodes increases), and fixed number of edges between each pair of layers (equal to the number of nodes). Each layer is given a number, and the interlayer networks are created between ordered pairs of layers, such that the interlayer networks are not symmetric with respect to the layers (node 1 on layer $a$ being connected to node 2 on layer $b$ doesn't imply node 2 on layer $a$ being connected to node 1 on layer $b$). Since between each pair of layers there is one geometric network, the mean interlayer degree of a node-layer is the number of layers minus one. We vary the number of elementary layers in the first aspect from 3 to 10 and the number of elementary layers in the second aspect from 1 to 4 in the 1000-node networks and from 1 to 2 in the other networks. Algorithm running times are presented in Figure \ref{fig:cpp_geo_absolute_scatter} as a function of the number of subnetworks found, which reflects the "size" or complexity of the enumeration problem. The running times of both \alg{nlse} and \alg{elsse} scale between $\mathcal{O}(n)$ and $\mathcal{O}(n^{1.2})$, where $n$ is the number of subnetworks found.

\begin{figure}[!t]
    \centering
    \includegraphics[width=0.6\columnwidth]{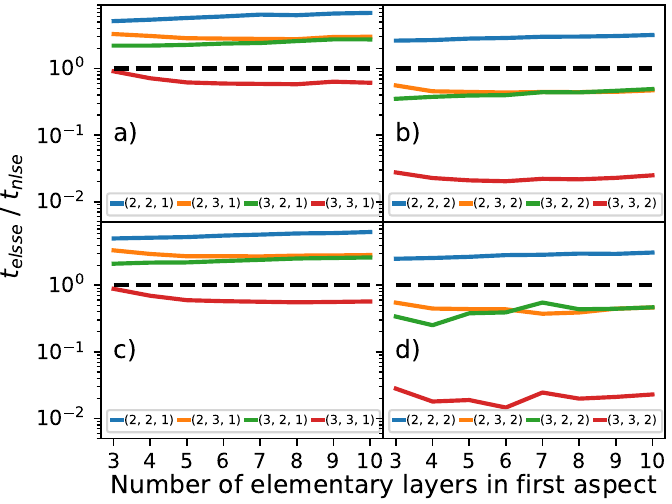}
    \caption{Relative running times as function of number of elementary layers in first aspect in two-aspect multilayer geometric random networks with intra- and interlayer networks generated from same node positions. Colors correspond to subnetwork sizes.
    a) 1000 nodes, 1500 intralayer edges, 1000 interlayer edges between each pair of layers, and 1 elementary layer in the second aspect.
    b) 1000 nodes, 1500 intralayer edges, 1000 interlayer edges between each pair of layers, and 2 elementary layers in the second aspect.
    c) 10 000 nodes, 15 000 intralayer edges, 10 000 interlayer edges between each pair of layers, and 1 elementary layer in the second aspect.
    d) 10 000 nodes, 15 000 intralayer edges, 10 000 interlayer edges between each pair of layers, and 2 elementary layers in the second aspect.
    }
    \label{fig:cpp_geo_mlayer}
\end{figure}

To visualize the exact causes of difference in the performance of \alg{nlse} and \alg{elsse}, the relative running times of both are shown in Figure \ref{fig:cpp_geo_mlayer} for networks of 1000 and 10 000 nodes and 1 and 2 layers in the second aspect for various subnetwork sizes. As the number of elementary layers in the first aspect increases from 3 to 10, the ratio of \alg{elsse}'s running time to \alg{nlse}'s running time stays nearly constant. Further, increasing the number of nodes from 1000 to 10 000 does not affect the ratio.
In contrast, increasing the number of elementary layers in the second aspect from 1 to 2 in the networks and enumerated subnetworks has the effect that \alg{elsse} is faster than \alg{nlse} on smaller number of nodes and elementary layers in the first aspect. This is due to an increase of the number of node-layers in the subnetworks (a subnetwork of size 2,2,1 has four node-layers and a subnetwork of size 2,2,2 has eight node-layers) and not to the increased number of elementary layers in the second aspect in the larger network, since the ratio of the running times stays the same for networks with 3 and 4 elementary layers in the second aspect and the same subnetwork sizes as the 2-elementary-layer case (see Supplementary Information S3).

\section{Discussion}
Enumerating subnetworks of application-agnostic, general-form multilayer networks is not a trivial task, and it is not computationally feasible with naive methods.
Both algorithms presented in this article, \alg{nlse} and \alg{elsse}, can be used for complete enumeration or unbiased sampling of connected induced subnetworks of multilayer networks.
They approach the enumeration problem differently, recursively expanding sets of node-layers or sets of elementary layers, respectively.

There are large differences between the performances of the two algorithms, and it is not easy to choose, a priori, which one to use on a given multilayer network and given subnetwork size.
In single-aspect multiplex networks, it seems that \alg{nlse} is generally faster than \alg{elsse} for small subnetwork sizes (fewer than 9 node-layers in total), based on experimental running times in multiplex Erd\H{o}s-Rényi networks (low edge overlap between layers), and multiplex geometric networks (high edge overlap between layers). In the geometric networks with high edge overlap, the threshold where \alg{elsse} becomes faster than \alg{nlse} is reached at smaller subnetwork size than in the Erd\H{o}s-Rényi networks with low edge overlap. However, in real-world multiplex protein-protein interaction networks with variable numbers of nodes and layers, \alg{nlse} is faster for almost all of the networks and subnetwork sizes up to 4 nodes and 3 layers (12 node-layers). The result is markedly different from the synthetic Erd\H{o}s-Rényi and geometric networks. Similarly to the single-aspect case, in two-aspect geometric multilayer networks with proper multilayer structure between layers \alg{nlse} is faster than \alg{elsse} for small subnetworks of size 2,2,2 (eight node-layers) but slower for subnetworks with more node-layers. Together, the results suggest that subnetwork size is the most significant factor in determining whether \alg{nlse} or \alg{elsse} is faster (with \alg{nlse} being faster for small subnetworks), but network structure also plays a significant role and affects where the relative performance transition point is in terms of subnetwork size. Since both algorithms seem to have domains where they outperform the other, and since distinguishing these domains from each other is not trivial, it is recommended to run both algorithms and use the result from the one that finishes first.

Subnetwork enumeration is important in many different network analysis methods, such as motif analysis and network comparison, and in many application areas, such as cellular biology and neuroscience. The algorithms \alg{nlse} and \alg{elsse} can be used "out-of-the-box" for multilayer subnetwork enumeration or sampling in most applications that are extended to the multilayer network realm.
This work aims to pave the way for deeper insights into network structure in multilayer networks, allowing researchers to uncover hidden patterns across interconnected layers in rich multilayered data. These tools will empower researchers to unravel the complexity in various domains, unlocking the potential of subnetwork analysis for increasingly complex interconnected systems with multiple modalities, data dimensions and several types of interactions.

\FloatBarrier

\bibliographystyle{ieeetr}
\bibliography{bibliography}

\section*{Acknowledgments}

Academy of Finland, 349366  Academy Research Fellow to M.K.

\pagebreak
\begin{center}
\textbf{\large Supplementary Information for Subnetwork enumeration algorithms for multilayer networks}
\end{center}

\section*{S1: Naive approaches to subnetwork enumeration}

In principle, one could use an existing subgraph enumeration algorithm to find multilayer subnetworks of a given size by enumerating a set of subgraphs of the underlying graph that is sure to contain all the sub\emph{networks} we are looking for (see Figure 1 in main article). However, in a naive approach, one would have to enumerate all subgraphs of the underlying graph with the number of nodes equal to the maximum number of node-layers possible in the span of the subnetworks. For example, when looking for subnetworks with 4 nodes, 2 layers in the first aspect, and 1 layer in the second aspect, one must enumerate all subgraphs of the underlying graph with up to $4 \times 2 \times 1 = 8$ node-layers when using an existing subgraph enumeration algorithm. There are two major problems with this approach: 1) the enumeration quickly becomes computationally infeasible as the number of elementary layers in the different aspects grows, and 2) most subgraphs of the underlying graph do not exactly correspond to a valid subnetwork defined in terms of sets of elementary layers, meaning that most of the work done in the enumeration is not fruitful. For example, consider a multilayer network of 1000 nodes, 5 elementary layers in the first aspect, and 5 elementary layers in the second aspect, and which has all the edges present (i.e., a complete network). There are ${1000 \choose 2} \times {5 \choose 2} \times {5 \choose 2} = 4.995 \times 10^7$ connected induced subnetworks of size $2,2,2$ in the network. To find all these subnetworks using the above mentioned subgraph of the underlying graph enumeration technique, we would have to enumerate all subgraphs of the underlying graph that contain $2 \times 2 \times 2 = 8$ node-layers, of which there are ${{1000 \times 5 \times 5} \choose {8}} \approx 3.78 \times 10^{30}$. Only approximately one in $1.32 \times 10^{23}$ of these subgraphs of the underlying graph therefore corresponds to a subnetwork, so this kind of enumeration method would do a massive amount of useless work.

Another, less naive approach would be to aggregate the multilayer network to one layer only, enumerate the subgraphs in that single-layer network, and then for each subgraph check all the layer combinations of desired size and whether the subnetwork corresponding to the nodes of the subgraph and the combination of layers is connected. If there is a high density of interlayer edges between layers, then many of the subgraphs and layer combinations correspond to a connected subnetwork. The key is the connectivity between layers which determines if an arbitrary combination of layers combined with a subgraph is likely to result in a connected subnetwork. If the interlayer density is low, most of the combinations of layers will not yield a connected subnetwork, and the rapid increase in possible combinations of layers as the number of layers and aspects grows quickly makes this approach computationally wasteful.

\section*{S2: Subnetworks in protein-protein interaction networks}

Numbers of subnetworks found and \alg{nlse} and \alg{elsse} running times and output rates are listed in Table \ref{table:number_of_subnets_ppi}.

\begin{table}[b]
    \tiny
    \centering
    \caption{Number of subnetworks of different sizes in the protein-protein interaction networks.}
    \begin{tabular}{@{\extracolsep{2pt}}c|cccccc}
 & & & \multicolumn{2}{c}{Running time (s)} & \multicolumn{2}{c}{Output rate (subnetworks/s)} \\
 \cline{4-5}
 \cline{6-7}
 & Subnetwork size & Number of subnetworks & nlse & elsse & nlse & elsse \\
 \hline
arabidopsis & 2, 2 & $1.40 \times 10^{4}$ & $1.84 \times 10^{0}$ & $5.06 \times 10^{0}$ & $7.60 \times 10^{3}$ & $2.77 \times 10^{3}$ \\
 & 3, 2 & $1.26 \times 10^{6}$ & $7.42 \times 10^{2}$ & $2.09 \times 10^{3}$ & $1.70 \times 10^{3}$ & $6.05 \times 10^{2}$ \\
 & 2, 3 & $5.02 \times 10^{3}$ & $4.40 \times 10^{0}$ & $1.06 \times 10^{1}$ & $1.14 \times 10^{3}$ & $4.74 \times 10^{2}$ \\
 & 3, 3 & $2.43 \times 10^{5}$ & $2.44 \times 10^{3}$ & $5.17 \times 10^{3}$ & $9.98 \times 10^{1}$ & $4.71 \times 10^{1}$ \\
 & 4, 2 & $4.35 \times 10^{8}$ & $3.23 \times 10^{5}$ & $9.23 \times 10^{5}$ & $1.35 \times 10^{3}$ & $4.72 \times 10^{2}$ \\
 & 4, 3 & $7.52 \times 10^{7}$ & $1.30 \times 10^{6}$ & $2.43 \times 10^{6}$ & $5.80 \times 10^{1}$ & $3.10 \times 10^{1}$ \\
bos & 2, 2 & $1.18 \times 10^{2}$ & $6.00 \times 10^{-3}$ & $1.53 \times 10^{-2}$ & $1.97 \times 10^{4}$ & $7.71 \times 10^{3}$ \\
 & 3, 2 & $4.21 \times 10^{2}$ & $2.55 \times 10^{-2}$ & $5.74 \times 10^{-2}$ & $1.65 \times 10^{4}$ & $7.34 \times 10^{3}$ \\
 & 2, 3 & $2.20 \times 10^{1}$ & $4.57 \times 10^{-3}$ & $1.24 \times 10^{-2}$ & $4.81 \times 10^{3}$ & $1.78 \times 10^{3}$ \\
 & 3, 3 & $1.20 \times 10^{2}$ & $2.20 \times 10^{-2}$ & $5.48 \times 10^{-2}$ & $5.45 \times 10^{3}$ & $2.19 \times 10^{3}$ \\
 & 4, 2 & $1.69 \times 10^{3}$ & $7.28 \times 10^{-2}$ & $2.43 \times 10^{-1}$ & $2.32 \times 10^{4}$ & $6.95 \times 10^{3}$ \\
 & 4, 3 & $5.46 \times 10^{2}$ & $1.26 \times 10^{-1}$ & $3.09 \times 10^{-1}$ & $4.34 \times 10^{3}$ & $1.77 \times 10^{3}$ \\
candida & 2, 2 & $1.32 \times 10^{2}$ & $5.16 \times 10^{-2}$ & $1.08 \times 10^{-1}$ & $2.56 \times 10^{3}$ & $1.22 \times 10^{3}$ \\
 & 3, 2 & $5.58 \times 10^{3}$ & $2.41 \times 10^{0}$ & $8.94 \times 10^{0}$ & $2.32 \times 10^{3}$ & $6.24 \times 10^{2}$ \\
 & 2, 3 & $3.00 \times 10^{1}$ & $2.61 \times 10^{-2}$ & $8.42 \times 10^{-2}$ & $1.15 \times 10^{3}$ & $3.56 \times 10^{2}$ \\
 & 3, 3 & $1.46 \times 10^{3}$ & $2.54 \times 10^{0}$ & $1.24 \times 10^{1}$ & $5.76 \times 10^{2}$ & $1.18 \times 10^{2}$ \\
 & 4, 2 & $5.49 \times 10^{5}$ & $1.68 \times 10^{2}$ & $8.89 \times 10^{2}$ & $3.26 \times 10^{3}$ & $6.18 \times 10^{2}$ \\
 & 4, 3 & $1.70 \times 10^{5}$ & $2.09 \times 10^{2}$ & $1.20 \times 10^{3}$ & $8.13 \times 10^{2}$ & $1.42 \times 10^{2}$ \\
celegans & 2, 2 & $6.96 \times 10^{3}$ & $7.52 \times 10^{-1}$ & $2.04 \times 10^{0}$ & $9.25 \times 10^{3}$ & $3.41 \times 10^{3}$ \\
 & 3, 2 & $5.46 \times 10^{5}$ & $6.62 \times 10^{1}$ & $2.07 \times 10^{2}$ & $8.24 \times 10^{3}$ & $2.63 \times 10^{3}$ \\
 & 2, 3 & $3.12 \times 10^{3}$ & $1.59 \times 10^{0}$ & $4.05 \times 10^{0}$ & $1.97 \times 10^{3}$ & $7.72 \times 10^{2}$ \\
 & 3, 3 & $3.22 \times 10^{5}$ & $2.67 \times 10^{2}$ & $6.73 \times 10^{2}$ & $1.21 \times 10^{3}$ & $4.78 \times 10^{2}$ \\
 & 4, 2 & $6.74 \times 10^{7}$ & $1.19 \times 10^{4}$ & $3.32 \times 10^{4}$ & $5.66 \times 10^{3}$ & $2.03 \times 10^{3}$ \\
 & 4, 3 & $4.42 \times 10^{7}$ & $5.03 \times 10^{4}$ & $1.19 \times 10^{5}$ & $8.79 \times 10^{2}$ & $3.73 \times 10^{2}$ \\
drosophila & 2, 2 & $5.54 \times 10^{4}$ & $2.08 \times 10^{0}$ & $6.51 \times 10^{0}$ & $2.66 \times 10^{4}$ & $8.50 \times 10^{3}$ \\
 & 3, 2 & $1.77 \times 10^{6}$ & $1.11 \times 10^{2}$ & $3.39 \times 10^{2}$ & $1.60 \times 10^{4}$ & $5.24 \times 10^{3}$ \\
 & 2, 3 & $3.60 \times 10^{4}$ & $6.05 \times 10^{0}$ & $1.56 \times 10^{1}$ & $5.96 \times 10^{3}$ & $2.31 \times 10^{3}$ \\
 & 3, 3 & $1.50 \times 10^{6}$ & $4.76 \times 10^{2}$ & $9.20 \times 10^{2}$ & $3.16 \times 10^{3}$ & $1.63 \times 10^{3}$ \\
 & 4, 2 & $7.93 \times 10^{7}$ & $7.39 \times 10^{3}$ & $1.83 \times 10^{4}$ & $1.07 \times 10^{4}$ & $4.32 \times 10^{3}$ \\
 & 4, 3 & $7.98 \times 10^{7}$ & $4.49 \times 10^{4}$ & $5.95 \times 10^{4}$ & $1.78 \times 10^{3}$ & $1.34 \times 10^{3}$ \\
gallus & 2, 2 & $1.29 \times 10^{2}$ & $2.76 \times 10^{-2}$ & $7.60 \times 10^{-2}$ & $4.67 \times 10^{3}$ & $1.70 \times 10^{3}$ \\
 & 3, 2 & $2.72 \times 10^{3}$ & $7.36 \times 10^{-1}$ & $2.12 \times 10^{0}$ & $3.70 \times 10^{3}$ & $1.28 \times 10^{3}$ \\
 & 2, 3 & $3.00 \times 10^{1}$ & $1.59 \times 10^{-2}$ & $4.79 \times 10^{-2}$ & $1.89 \times 10^{3}$ & $6.26 \times 10^{2}$ \\
 & 3, 3 & $9.71 \times 10^{2}$ & $9.14 \times 10^{-1}$ & $2.99 \times 10^{0}$ & $1.06 \times 10^{3}$ & $3.25 \times 10^{2}$ \\
 & 4, 2 & $1.28 \times 10^{5}$ & $1.74 \times 10^{1}$ & $8.41 \times 10^{1}$ & $7.36 \times 10^{3}$ & $1.52 \times 10^{3}$ \\
 & 4, 3 & $4.79 \times 10^{4}$ & $2.57 \times 10^{1}$ & $1.24 \times 10^{2}$ & $1.86 \times 10^{3}$ & $3.86 \times 10^{2}$ \\
mus & 2, 2 & $2.52 \times 10^{4}$ & $1.32 \times 10^{0}$ & $3.51 \times 10^{0}$ & $1.92 \times 10^{4}$ & $7.18 \times 10^{3}$ \\
 & 3, 2 & $1.17 \times 10^{6}$ & $1.14 \times 10^{2}$ & $3.38 \times 10^{2}$ & $1.03 \times 10^{4}$ & $3.46 \times 10^{3}$ \\
 & 2, 3 & $1.65 \times 10^{4}$ & $3.23 \times 10^{0}$ & $8.67 \times 10^{0}$ & $5.11 \times 10^{3}$ & $1.90 \times 10^{3}$ \\
 & 3, 3 & $8.94 \times 10^{5}$ & $3.96 \times 10^{2}$ & $1.01 \times 10^{3}$ & $2.26 \times 10^{3}$ & $8.84 \times 10^{2}$ \\
 & 4, 2 & $9.32 \times 10^{7}$ & $1.12 \times 10^{4}$ & $3.44 \times 10^{4}$ & $8.33 \times 10^{3}$ & $2.71 \times 10^{3}$ \\
 & 4, 3 & $7.93 \times 10^{7}$ & $5.36 \times 10^{4}$ & $1.11 \times 10^{5}$ & $1.48 \times 10^{3}$ & $7.13 \times 10^{2}$ \\
plasmodium & 2, 2 & $3.30 \times 10^{1}$ & $5.37 \times 10^{-2}$ & $1.91 \times 10^{-1}$ & $6.15 \times 10^{2}$ & $1.72 \times 10^{2}$ \\
 & 3, 2 & $7.01 \times 10^{2}$ & $8.38 \times 10^{-1}$ & $3.15 \times 10^{0}$ & $8.37 \times 10^{2}$ & $2.23 \times 10^{2}$ \\
 & 2, 3 & 0 & $3.18 \times 10^{-2}$ & $1.12 \times 10^{-1}$ & - & - \\
 & 3, 3 & 0 & $7.68 \times 10^{-1}$ & $2.93 \times 10^{0}$ & - & - \\
 & 4, 2 & $1.74 \times 10^{4}$ & $1.90 \times 10^{1}$ & $8.89 \times 10^{1}$ & $9.18 \times 10^{2}$ & $1.96 \times 10^{2}$ \\
 & 4, 3 & 1 & $1.93 \times 10^{1}$ & $9.05 \times 10^{1}$ & $5.18 \times 10^{-2}$ & $1.11 \times 10^{-2}$ \\
rattus & 2, 2 & $4.54 \times 10^{3}$ & $6.67 \times 10^{-1}$ & $1.65 \times 10^{0}$ & $6.82 \times 10^{3}$ & $2.75 \times 10^{3}$ \\
 & 3, 2 & $7.39 \times 10^{5}$ & $1.88 \times 10^{2}$ & $5.92 \times 10^{2}$ & $3.93 \times 10^{3}$ & $1.25 \times 10^{3}$ \\
 & 2, 3 & $1.56 \times 10^{3}$ & $1.46 \times 10^{0}$ & $4.63 \times 10^{0}$ & $1.07 \times 10^{3}$ & $3.37 \times 10^{2}$ \\
 & 3, 3 & $3.76 \times 10^{5}$ & $5.66 \times 10^{2}$ & $1.95 \times 10^{3}$ & $6.64 \times 10^{2}$ & $1.93 \times 10^{2}$ \\
 & 4, 2 & $1.90 \times 10^{8}$ & $5.13 \times 10^{4}$ & $1.79 \times 10^{5}$ & $3.70 \times 10^{3}$ & $1.06 \times 10^{3}$ \\
 & 4, 3 & $1.03 \times 10^{8}$ & $1.51 \times 10^{5}$ & $5.55 \times 10^{5}$ & $6.80 \times 10^{2}$ & $1.85 \times 10^{2}$ \\
sacchcere & 2, 2 & $1.25 \times 10^{6}$ & $1.92 \times 10^{2}$ & $4.36 \times 10^{2}$ & $6.53 \times 10^{3}$ & $2.88 \times 10^{3}$ \\
 & 3, 2 & $2.06 \times 10^{8}$ & $8.91 \times 10^{4}$ & $1.33 \times 10^{5}$ & $2.31 \times 10^{3}$ & $1.55 \times 10^{3}$ \\
 & 2, 3 & $2.65 \times 10^{6}$ & $1.06 \times 10^{3}$ & $1.57 \times 10^{3}$ & $2.50 \times 10^{3}$ & $1.69 \times 10^{3}$ \\
 & 3, 3 & $5.30 \times 10^{8}$ & $1.15 \times 10^{6}$ & $8.33 \times 10^{5}$ & $4.59 \times 10^{2}$ & $6.36 \times 10^{2}$ \\
 & 4, 2 & - & - & - & - & - \\
 & 4, 3 & - & - & - & - & - \\
sacchpomb & 2, 2 & $2.01 \times 10^{5}$ & $1.89 \times 10^{1}$ & $5.41 \times 10^{1}$ & $1.06 \times 10^{4}$ & $3.71 \times 10^{3}$ \\
 & 3, 2 & $2.15 \times 10^{7}$ & $3.47 \times 10^{3}$ & $8.09 \times 10^{3}$ & $6.20 \times 10^{3}$ & $2.66 \times 10^{3}$ \\
 & 2, 3 & $2.94 \times 10^{5}$ & $8.36 \times 10^{1}$ & $1.78 \times 10^{2}$ & $3.51 \times 10^{3}$ & $1.65 \times 10^{3}$ \\
 & 3, 3 & $3.80 \times 10^{7}$ & $2.64 \times 10^{4}$ & $3.87 \times 10^{4}$ & $1.44 \times 10^{3}$ & $9.82 \times 10^{2}$ \\
 & 4, 2 & $3.48 \times 10^{9}$ & $8.36 \times 10^{5}$ & $1.60 \times 10^{6}$ & $4.16 \times 10^{3}$ & $2.18 \times 10^{3}$ \\
 & 4, 3 & - & - & - & - & - \\
    \end{tabular}
    \label{table:number_of_subnets_ppi}
\end{table}

\section*{S3: More results from two-aspect geometric random multilayer networks}

\begin{figure}[t]
    \centering
    \begin{subfigure}[t]{0.49\columnwidth}
        \includegraphics[width=\columnwidth]{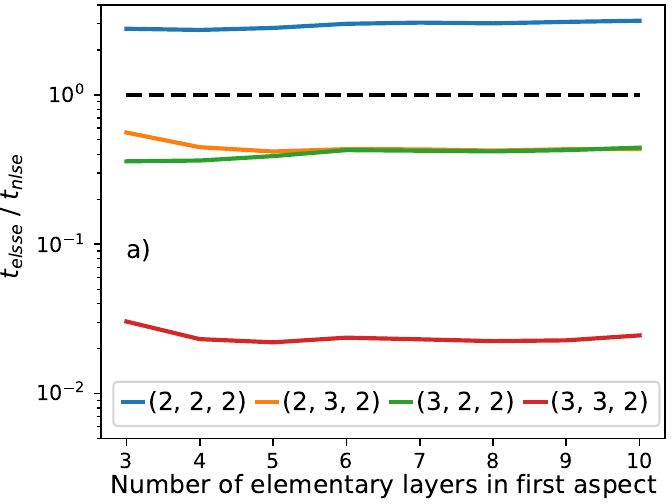}
    \end{subfigure}
    \begin{subfigure}[t]{0.49\columnwidth}
        \includegraphics[width=\columnwidth]{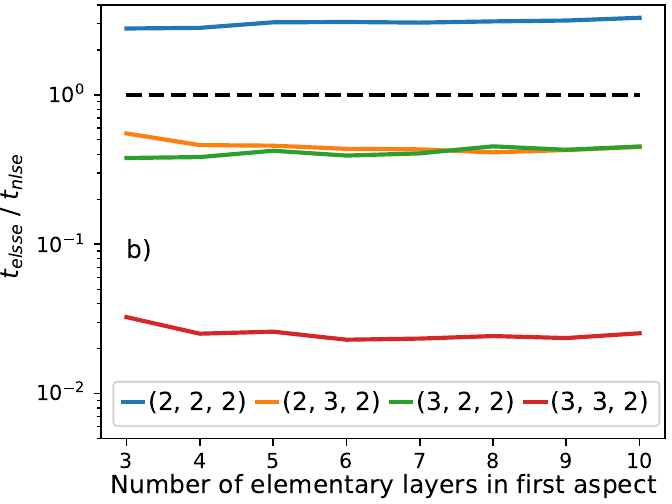}
    \end{subfigure}
    \caption{Relative running times in two-aspect multilayer geometric random networks as a function of the number of elementary layers in the first aspect with a) three and b) four elementary layers in the second aspect (1000 nodes, 1500 intralayer edges per layer, 1000 interlayer edges between each pair of layers). Colors correspond to subnetwork sizes. The figures are nearly identical: subnetwork size heavily influences the relative running times, with \alg{elsse} performing better for larger subnetworks, and the number of layers in the first or second aspect in the multilayer network affects the relative running times only very little, if at all.}
    \label{fig:cpp_geo_3_and_4}
\end{figure}

Relative running times (\alg{elsse} divided by \alg{nlse}) for two-aspect multilayer geometric random networks with three and four elementary layers in the second aspect are shown in Figure \ref{fig:cpp_geo_3_and_4}. They are nearly identical to each other and to the case with two elementary layers in the second aspect (see main article). Therefore, the main determining factor of the relative performance is subnetwork size.

\end{document}